\documentclass[
 aps,
 amsmath,amssymb,
 reprint,
]{revtex4-1}

\usepackage{graphicx}
\usepackage{dcolumn}
\usepackage{bm}
\usepackage[utf8]{inputenc}
\usepackage[T1]{fontenc}
\usepackage{mathptmx}
\usepackage{etoolbox}

\makeatletter
\def\affiliation#1{\begingroup\def\@affiliation{#1}\endgroup}
\makeatother

\makeatletter
\def\@email#1#2{%
 \endgroup
 \patchcmd{\titleblock@produce}
  {\frontmatter@RRAPformat}
  {\frontmatter@RRAPformat{\produce@RRAP{*#1\href{mailto:#2}{#2}}}\frontmatter@RRAPformat}
  {}{}
}
\makeatother
\makeatother
\begin{document}


\title
{Layer breathing Raman mode in two-dimensional van der Waals material $\mathrm{Cr_2Ge_2Te_6}$\\}

\author{Nilesh Choudhury}
\author{Neesha Yadav}
\author{Sandeep}
\author{Mayank Shukla}
\author{Pintu Das}%
\affiliation{ 
Physics Department, Indian Institute of Technology Delhi, New Delhi, 110016, India
}%

\date{\today}

\begin{abstract}
 Two-dimensional (2D) van der Waals (vdW) magnetic materials have emerged as key materials for next-generation magneto-electric and spintronic devices, where understanding the relationship between layer number, lattice dynamics, and magnetic interactions is very important. In this work, we report the observation of the layer breathing mode (LBM) in few-layer $\mathrm{Cr_2Ge_2Te_6}$, a ferromagnetic semiconductor with thickness dependent electronic, magnetic and optical properties, using Raman spectroscopy, which serves as a direct fingerprint of interlayer coupling and lattice symmetry. Group-theoretical symmetry analysis confirms that the CGT falls under the non-polar category of layered material. The evolution of the LBM-frequency with increasing layer number (N) reveals a distinct softening trend, characteristic of weakening restoring forces in thicker flakes. By fitting the experimental Raman data using the Linear Chain Model (LCM), we quantitatively extract the interlayer force constant ($\mathrm{K_c}$), providing a measure of the vdW coupling strength between layers. 
\end{abstract}

\maketitle

\section{\label{sec:level1} Introduction \\}

Two-dimensional (2D) materials are a broad category of materials in which the atomic layers are covalently or ionically bonded within the plane, while adjacent layers are weakly coupled through van der Waals (vdW) forces \cite{ningrum2020recent}. There are more than 130 known thermodynamically stable 2D materials among which transition metal dichalcogenides (TMDCs), hexagonal boron nitride (h-BN), graphene, and phosphorene are the most popular ones due to their unique optical, electronic and magnetic properties \cite{shrivastava}. The characteristic of 2D materials arises from the interplay between the intrinsic characteristics of each individual layer, the general structural symmetry, and the strength of the interaction between adjacent layers \cite{liang}. \\
As 2D vdW material is thinned down to a few or multiple layers, its optical, electronic and magnetic characteristics change markedly with the number of layers, primarily because of variations in interlayer interactions \cite{bonaccorso2013multiwall}. To fully exploit these tunable properties, it is important to have reliable methods to identify the exact layer thickness in fabricated structures or devices. While optical contrast techniques such as atomic force microscopy (AFM) and optical microscopy are commonly used for this purpose, their accuracy often depends on substrate conditions and illumination parameters \cite{puretzky2015low}. Alternatively, Raman and infrared (IR) spectroscopies serve as more robust and informative approaches, as they directly probe lattice vibrations. These spectroscopic methods not only allow precise determination of layer number but also provide deeper insights into magnetic and electronic characteristics, strain, doping, and interlayer interactions in 2D materials \cite{li2017layer}. \\
\begin{figure}
	\centering
	\includegraphics[width=1\linewidth]{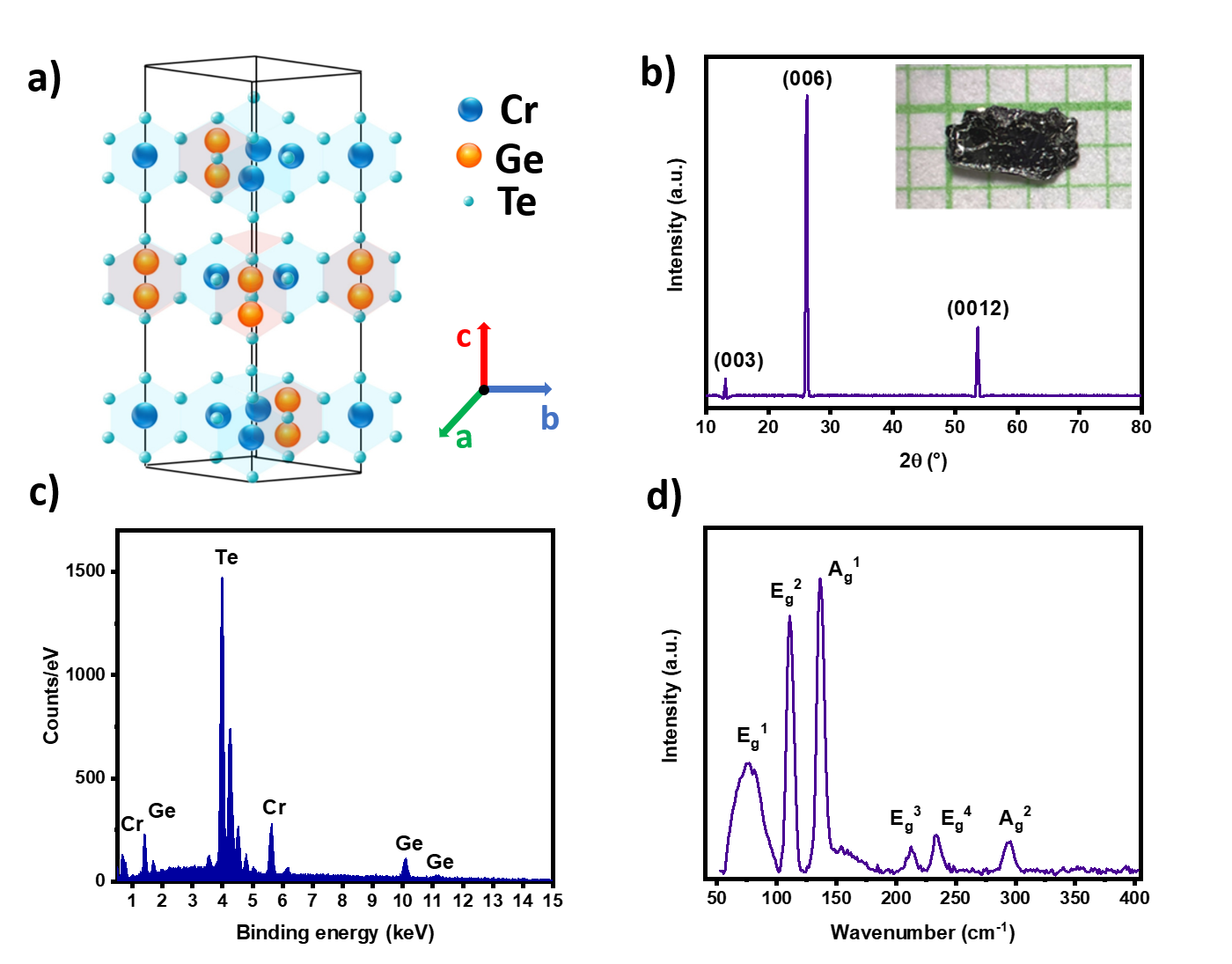}
	\caption{\label{fig:frog} a) The schematic of the crystal structure of CGT, b) XRD pattern of the bulk CGT crystal with inset showing an optical image of the crystal of size 3×2 mm, c) EDAX spectra for elemental confirmation. d) Raman spectra of bulk CGT crystal having out of plane Eg and in plane Ag vibrational modes.}
\end{figure}
2D vdW materials exhibit two fundamentally distinct categories of Raman-active vibrational modes \cite{pizzi}. The first category involves intralayer vibrations, where atoms within an individual layer move relative to each other, typically governed by strong covalent or ionic bonding. The second category corresponds to interlayer vibrations, in which each layer oscillates as a rigid unit relative to adjacent layers, giving rise to in-plane shear (S) and out of plane layer-breathing (LB) modes that are primarily determined by the weak vdW coupling between layers \cite{pizzi}. Interlayer LB and S modes have been extensively investigated through Raman spectroscopy across a wide range of layered materials. These modes have been reported in bulk systems such as graphite \cite{liu2022,lui2012,cong2014,tan2012shear,lui2013measurement,he2013observation}, hexagonal boron nitride (h-BN) \cite{stenger2017low},$\mathrm{MoS_2}$ \cite{boukhicha2013anharmonic}, $\mathrm{WS_2}$ \cite{zhao2013interlayer}, $\mathrm{NbSe_2}$ \cite{he2016interlayer}, $\mathrm{ReSe_2}$ \cite{zhao2015interlayer}, $\mathrm{Bi_2X_3}$ (X=Te and Se) \cite{zhao2014interlayer}, $\mathrm{MoTe_2}$\cite{song2016physical} etc. Since these interlayer Raman modes provide direct information about the interlayer coupling strength, understanding them is essential for thickness-dependent studies of vdW materials, where the strength of interlayer interactions governs the evolution of their structural, electronic, and magnetic properties.

Among the 2D vdW materials, $\mathrm{Cr_2Ge_2Te_6}$ (CGT) is a layered ferromagnetic semiconductor belonging to the family of 2D van der Waals (vdW) materials that has recently gained significant attention for its potential in spintronics and magneto-optical applications. The magnetic behavior of CGT in both bulk and few-layer forms can be well described within a Heisenberg-type spin model, suggesting dominant intralayer ferromagnetic exchange interactions modulated by interlayer coupling \cite{gong2017discovery}. Studying the Raman-active phonon modes of CGT is therefore crucial for understanding the fundamental coupling between lattice vibrations and magnetism in 2D ferromagnets. Such insights not only clarify how the magnetic anisotropy and Curie temperature evolve with the number of layers, but also offer valuable pathways to engineer magneto-elastic coupling and spin–phonon interactions for advanced device applications. In spintronic and nanoscale magneto-optical devices, where efficient spin, charge, and heat transport are essential, controlling phonon dynamics through strain, temperature, or external fields can offer a means to tune magnetic ordering and energy dissipation. Although intralayer phonon modes in CGT have been extensively studied through a combination of theoretical \cite{zhang2019first} and experimental \cite{tian2016magneto} Raman investigations, the interlayer vibrational modes have not yet been directly probed using optical techniques. The variation of intralayer phonon vibrational modes with thickness has previously been investigated, revealing that the intensity of Raman-active modes increases significantly as the number of layers decreases \cite{tian2016magneto}. \\
In this work, we performed a symmetry analysis of the CGT crystal structure to classify its layered nature and understand the evolution of its vibrational modes as a function of the number of layers. The experimentally observed interlayer LB mode frequencies in few-layer CGT show a systematic shift toward lower frequencies as the number of layers decreases. This behavior can be well described using a simple linear chain model (LCM) that considers only nearest-neighbor interactions, treating each layer as a single mass unit \cite{kittel2018introduction}. On the basis of this model, the interlayer force constant was extracted from the experimental data, providing insights into the interlayer coupling strength in CGT.

\section{Experimental details \\}

Single crystals of CGT were grown using the conventional self-flux method with high-purity chromium(Cr), germanium(Ge), and tellurium(Te) chunks. The elements were weighed in the appropriate stoichiometric ratio and inserted into an alumina crucible inside an argon-filled glovebox to minimize oxidation. The crucible was then sealed in an evacuated quartz ampoule which was subsequently placed in a heating furnace. The furnace was initially heated to 1020 $\mathrm{^oC}$ at a rate of 114 $\mathrm{^oC/hour}$ and kept at 1020 $\mathrm{^oC}$ for 24 hours to ensure complete homogenization, followed by cooling to 450 $\mathrm{^oC/hour}$. After the heat treatment, excess flux was removed using a centrifuge to obtain high-quality single crystals of CGT.\\
The grown crystals were characterized using X-ray diffraction(XRD), energy-dispersive X-ray spectroscopy (EDAX), and Raman spectroscopy to confirm their structural integrity, elemental composition, and corresponding phonon vibrational modes, respectively.\\
Thin flakes of CGT were mechanically exfoliated from the synthesized bulk single crystal using a scotch tape and subsequently dry transferred to a $\mathrm{Si/SiO_2}$ as substrate. The flakes were initially identified using an optical microscope based on their contrast, followed by determination of the thickness of the respective flakes by the AFM technique. Raman spectroscopy measurements were performed on the flakes at room temperature in a backscattering geometry using a solid state LASER of wavelength 523 nm and an excitation power of 23 mW. The backscattered signal was collected through a 50X objective lens and analyzed with a grating for spectral analysis.
\begin{figure}
	\centering
	\includegraphics[width=1\linewidth]{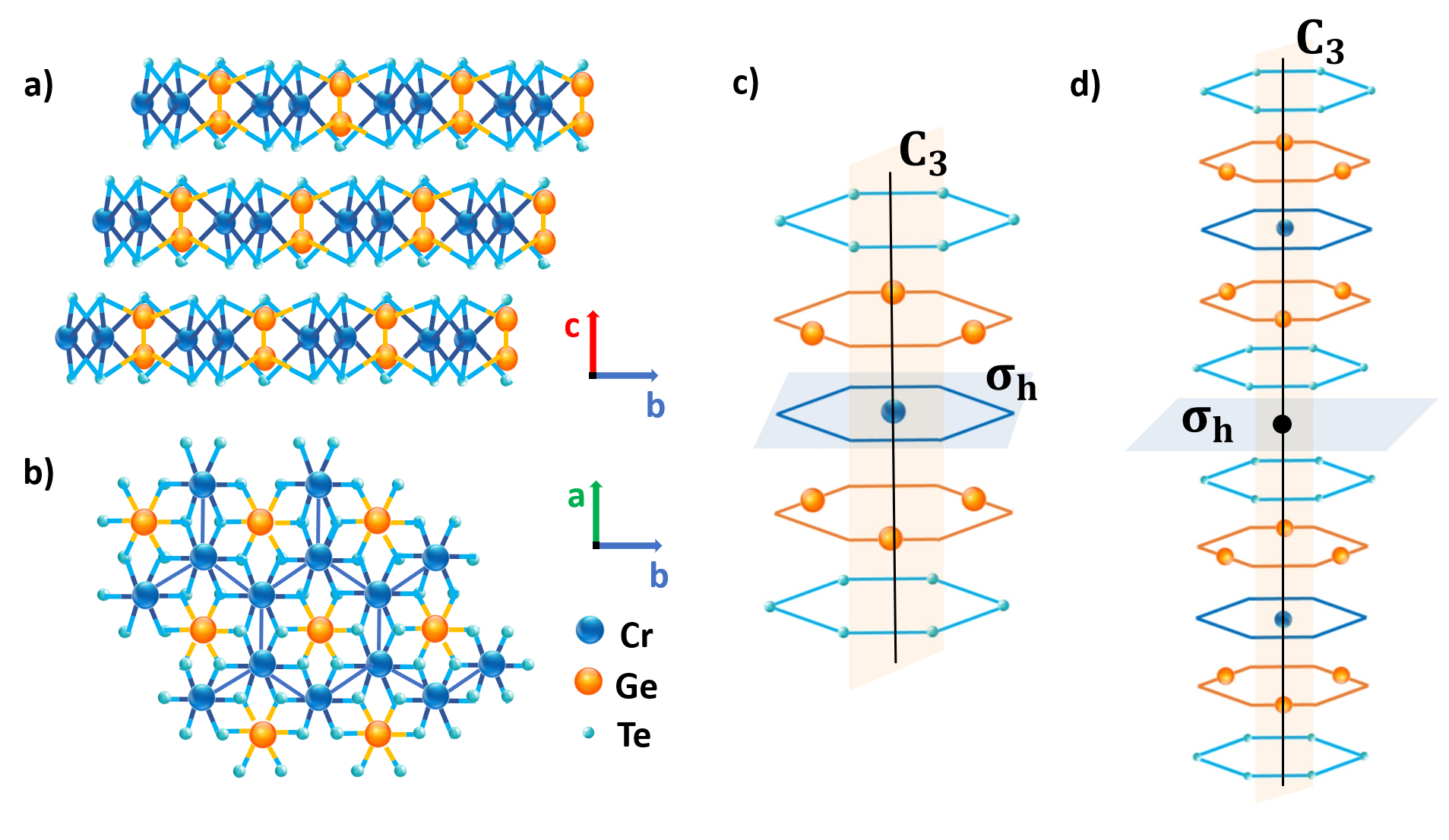}
	\caption{\label{fig:frog} Symmetry operations in layered CGT: a) schematic side view illustrating different layer arrangement within a single unit cell of CGT. b) top view of CGT crystal structure, showing the in-plane atomic arrangement. c) side view of monolayer showing with stacking sequence of Te-Ge-Cr-Ge-Te respectively. The black line indicates the threefold ($\mathrm{C_3}$) rotational axis, whereas the horizontal ($\mathrm{\sigma_h}$) and the vertical ($\mathrm{\sigma_v}$) reflection operations are shown as orange and grey planes, respectively. d) side view of two stacked layers of CGT, with an inversion center located within the horizontal ($\mathrm{\sigma_h}$) reflection plane.}
\end{figure}
\section{Results and discussion \\}

CGT is a layered TMDC that crystallizes in the rhombohedral $\mathrm{R3}$ space group (Hall No. 148), having trigonal symmetry \cite{sun2018effects}. Each unit cell of CGT contains three layers stacked along the (001) direction in an ABC stacking sequence as shown in Fig. 1(a) \cite{ji2013ferromagnetic}. The structure consists of slightly distorted Te octahedra, where $\frac{2}{3}$ of the octahedral sites are occupied by $\mathrm{Cr^{-3}}$ atoms, with uniform Cr-Te bond lengths of approxiamtely $\mathrm{2.79\AA}$, and the remaining $\frac{1}{3}$ by Ge-Ge dimers, with Ge-Te bond lengths of $\mathrm{2.60}$ \cite{carteaux1995crystallographic}.  
To confirm the crystalline structural properties of CGT, XRD was performed. Fig. 1(b) represents the XRD pattern of the single crystal CGT, with well-indexed diffraction peaks that confirm the expected hexagonal crystal structure with only (003), (006) and (0012) reflection planes observed, which is characteristic of a high-quality single crystal oriented along the c-axis. To confirm the elemental composition of the single crystal EDAX was performed as shown in Fig. 1(c). The atomic percentage ratio of the compositional elements was found to be Cr: Ge: Te = 1: 1.186: 3.155.
\\

Raman spectroscopy measurements were performed on the single crystal to characterize the normal phonon vibrational modes. The phonon modes at the center of the Brillouin zone can be described by the following irreducible representation,
\\
\[
    \mathrm{\Gamma= 5A_g + 5A_u+5E_g^1+5E_u^1+5E_g^2+5E_u^2}
\]
\\
where the $\mathrm{A_g}$,$\mathrm{E^1_g}$, and $\mathrm{E^2_g}$ modes are Raman active. Here, the $A_g$ mode is nondegenerate, while the $\mathrm{E^1_g}$ and $\mathrm{E^2_g}$ modes are doubly degenerate. The infrared-active modes possess $\mathrm{A_u}$, $\mathrm{E_u1}$, and $\mathrm{E_u^2}$ symmetry \cite{carteaux1995crystallographic}. Figure 1(d) displays the Raman spectrum of bulk crystal of CGT. Six distinct Raman-active modes are observed at 76.69 $\mathrm{cm^{-1}}$, 107.1 $\mathrm{cm^{-1}}$, 132.1 $\mathrm{cm^{-1}}$, 209.89 $\mathrm{cm^{-1}}$, 229.6 $\mathrm{cm^{-1}}$, and 292.5 $\mathrm{cm^{-1}}$. These spectral characteristics are consistent with both theoretical predictions and previous experimental reports \cite{zhang2019first}. Based on earlier studies, these modes correspond to the vibrational modes $\mathrm{E_g^1}$, $\mathrm{E_g^2}$, $\mathrm{A_g^1}$, $\mathrm{E_g^3}$, $\mathrm{E_g^4}$ and $\mathrm{A_g^2}$, respectively \cite{tian2016magneto}. Low-frequency phonon modes (below 150 $\mathrm{cm^{-1}}$) are predominantly contributed by Te atoms, because of their relatively large atomic mass. Vibration modes in the intermediate frequency range(180-240) $\mathrm{cm^{-1}}$) arise mainly from Cr atom displacements, while the high-frequency modes (>260 $\mathrm{cm^{-1}}$) are primarily associated with Ge atom vibrations \cite{zhang2019first}.
\\

\begin{figure}[h!]
\centering
\includegraphics[width=1\linewidth]{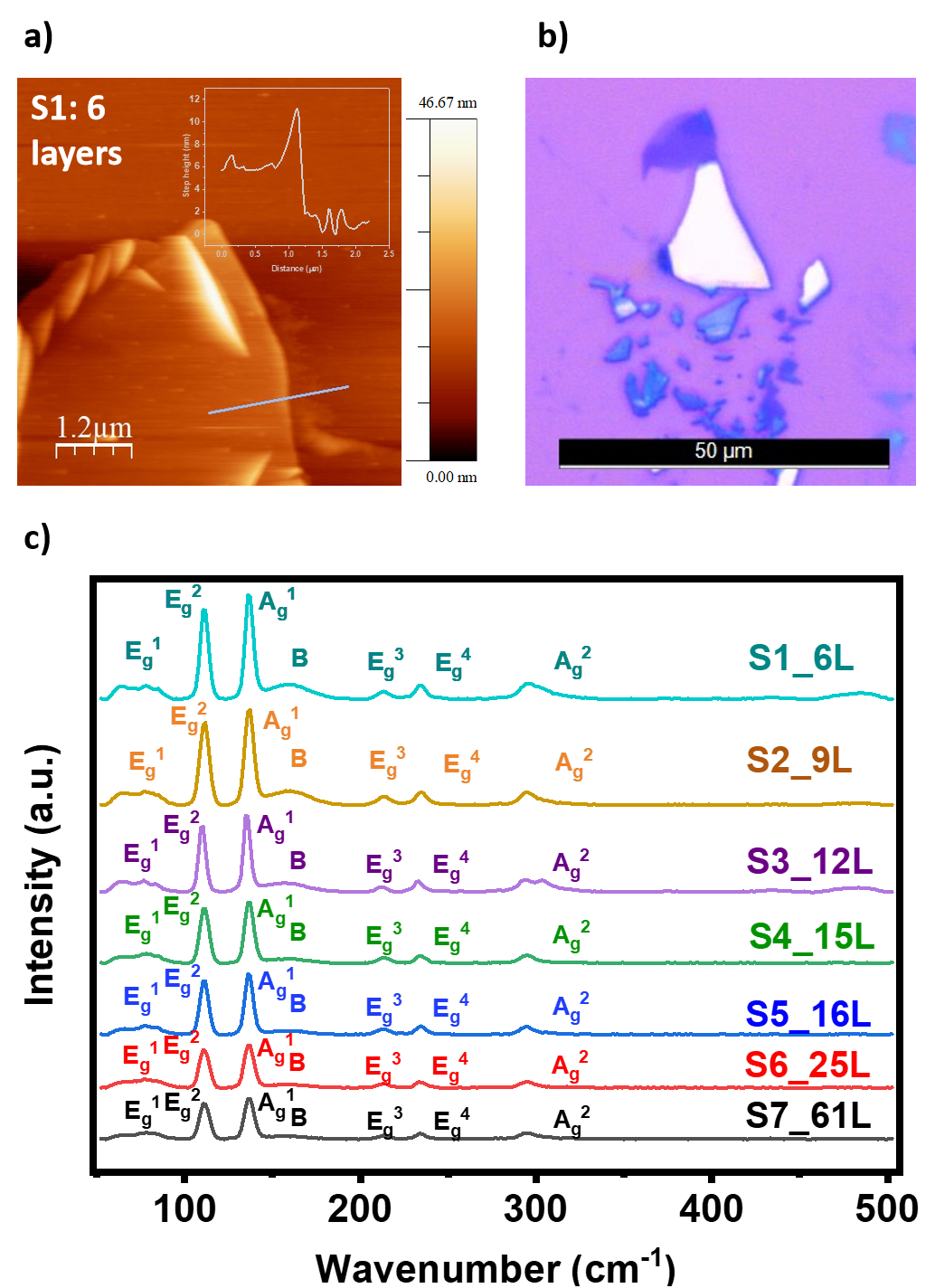}
\caption{\label{fig:frog} a) Atomic force microscopy (AFM) image of a six layer CGT flake on $Si/SiO_2$ substrate, the step height profile of the flake is shown as an inset. The corresponding optical image is also shown. b) Raman spectra of CGT flakes with varying thickness at room temperature, showing all the six phonon vibrational modes, with a distinct evolution in their intensities as a function of layer number.}

\end{figure}

In CGT, each layer follows the stacking sequence of Te-Ge-Cr-Ge-Te, forming a centrosymmetric, non-polar structure. This configuration preserves inversion symmetry within the layer so that, a symmetry operation that maps the layer onto itself when flipped upside down. The symmetry operation in such monolayer is called $\mathrm{\lambda-\rho}$ and multilayer  $\mathrm{\sigma-\rho}$, respectively \cite{pizzi}. In fig 2 (a) such stacking of one unit cell of CGT can be seen. The symmetry operations in mono and bi-layer CGT are represented in Fig 2(c) and (d), respectively. These symmetry operations for monolayer include: the identity operation (E), two threefold rotations ($\mathrm{C_3}$) with axes corresponding to clockwise and counterclockwise rotations, three  two-fold rotational axes in lying within the horizontal mirror plane ($\mathrm{\sigma_h}$), the horizontal mirror plane ($\mathrm{\sigma_h}$) and three vertical mirror planes ($\mathrm{\sigma_v}$) represented by the orange plane in Fig 2(c), three $\mathrm{S_3}$, screw axes leading to a $C_3$ followed by a reflection about ($\mathrm{\sigma_h}$) plane. For bilayer stacking, symmetry operations are identical to the monolayer, except that the horizontal mirror plane ($\mathrm{\sigma_h}$) is replaced by an inversion center. Because of the presence of an inversion center between two subsequent layers, CGT falls under the non-polar category of layered material. In non-polar layered structures, all adjacent layer pairs are symmetry-equivalent, enabling the use of a single interlayer force-constant matrix to describe the interactions between any neighboring layers \cite{pizzi}. 

Flakes of different thicknessess, labelled as S1-S7, were exfoliated from the CGT crystal via mechanical exfoliation. Fig. 3(a) shows the optical image of flake S1 alongside its AFM image. The inset of the AFM image corresponds to the height profile of the flake, confirming the presence of six layers. The layer numbers of such different flakes (S2-S7) were determined using the similar method (see supplementary Fig. S1). Raman spectroscopy measurements were carried out on the CGT flakes (S1-S7) with systematically varied thicknesses, as determined from the AFM analysis. The room temperature Raman spectra clearly display all six characteristic phonon vibrational modes of CGT as shown in Fig. 3(b). A pronounced evolution in the relative intensities and features of these modes is observed as the number of layers decreases from bulk to a few layers (see supplementary Fig. S2) , 

\begin{figure*}
\centering
\includegraphics[width=1\linewidth]{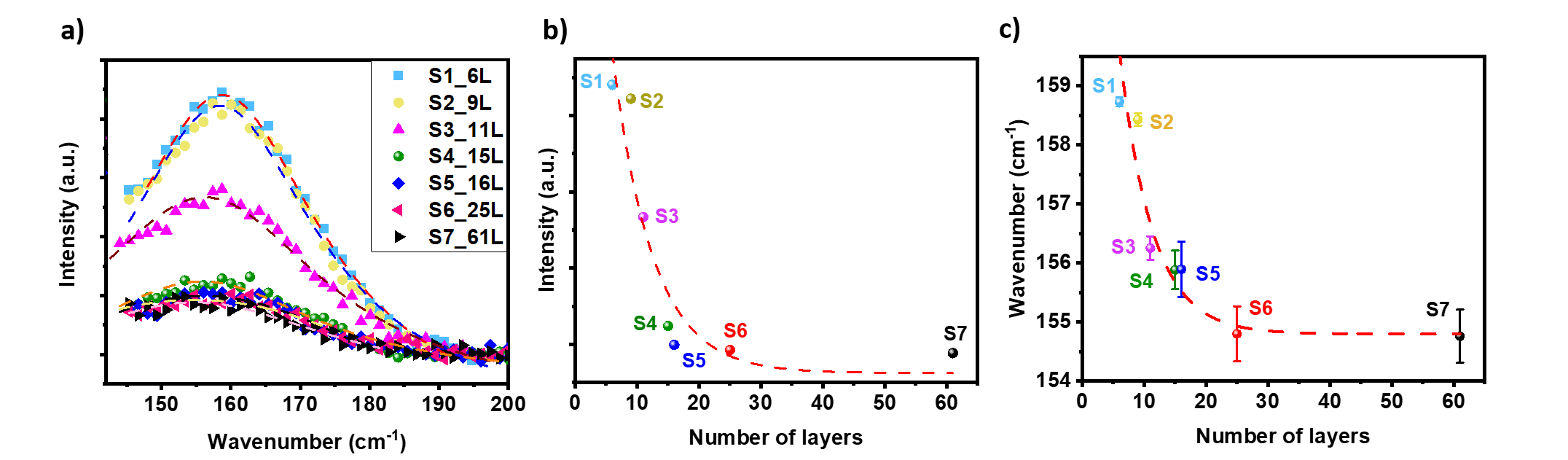}
\caption{\label{fig:frog} a) The scattered points are experimental data, while the dashed curves are Lorentzian fittings to the data for all the samples (S1-S7). b) change in the intensity as a function of number of layers for the layer breathing mode B (dashed line is guide to the eye), c) frequency evolution of breathing modes as a function of number of layers (dashed line is guide to the eye).}
\end{figure*}

reflecting the thickness-dependent vibrational properties of the material \cite{tian2016magneto}. In addition to the intralayer Raman-active modes, a distinct peak is observed around 158.73 $\mathrm{cm^{-1}}$, referred to as the B mode. The intensity of this mode increases significantly with decreasing layer number, indicating its sensitivity to the sample thickness. Such behavior is characteristic of an out-of-plane vibrational mode, commonly associated with interlayer coupling in layered van der Waals systems. The pronounced enhancement of the B mode in thinner flakes suggests a strong modulation of interlayer dynamics, highlighting its potential role as a spectroscopic fingerprint to probe the layer-dependent lattice interactions in CGT.\\
To accurately determine the intensity and peak position of the B peak, a Lorentzian fitting was performed on the Raman spectrum, as shown in Fig. 4(a). The detailed lineshape analysis of the spectra yielded excellent agreement with the Lorentzian profile. The extracted peak intensity and position of the B mode are presented as a function of the number of layers in Fig. 4(b) and 4(c). In addition to the observed intensity evolution, mode B exhibits a systematic frequency shift towards the longer wavelength with decreasing thickness of the flake as observed in the literature for different TMDCs, which is characteristic of a layer breathing Raman mode \cite{pizzi}. As more layers are added, the breathing mode becomes less stiff and, the energy associated with the vibration will decrease resulting in shifting of the mode towards the longer wavelength. Based on the characteristic behavior of this peak—its out-of-plane vibrational nature and strong dependence on the number of layers—the B peak can be identified as the layer breathing mode in CGT. The presence of an interlayer Raman mode in CGT has been further verified using data from Materials Cloud\cite{talirz2020materials}, an open-access platform for computational materials science, where the fan diagram clearly indicates the existence of such a mode around 156 $\mathrm{cm^{-1}}$ for N=6. \\
The layered material system is modeled as a finite linear chain of masses connected by springs with an interlayer force constant $\mathrm{K_c}$ as shown in the Fig. 5(a), which may vary depending on the direction of vibration. This model successfully captures the qualitative dependence of the frequency of the vibrational mode on the number of layers\cite{luo1996theory,michel2012theory}. The model considers only interlayer interactions of the nearest-neighbor as the dominant coupling mechanism, while the influence of the substrate is neglected for simplicity \cite{zhao2013interlayer}.
The breathing mode frequency, $\mathrm{v_B}$ can be expressed as,
\[
    \mathrm{v_{B}= \frac{1}{\sqrt{2}\pi c}\sqrt{\frac{K_c}{\mu}}\sqrt{1-cos{\frac{\pi}{N}}}} 
\]
The equation can be expressed as,
\[
    \mathrm{v_{B}= \frac{1}{\sqrt{2}\pi c}\sqrt{\frac{K_c}{\mu}}\sin{\frac{\pi}{2N}}} 
\]

where $\mathrm{\mu}$ is the reduced layer mass per unit area and N is the number of layers. To quantitatively analyze the dynamics of the interlayer, a simplified linear chain model is used to describe the vibrational modes between layers, where each structural unit consisting of a Te–Gr–Cr–Gr–Te layer is treated as a single rigid mass, as illustrated in Fig. 5(b). The layer mass per unit area, $\mathrm{\mu}$ for CGT is $\mathrm{4.08\times10^{-6} {Kg}/{m^{2}}}$ \cite{talirz2020materials}. To calculate the force constant between layers , the evolution of the phonon frequency of the LB mode with the number of layers has been fitted using the linear chain model. The experimental data are in excellent agreement with the LCM fit, as illustrated in Fig. 5(c). From the fit, the  calculated value of $\mathrm{K_c}$ is found to be $\mathrm{1.33 \pm 0.09\times10^{19} {Kg}/{m^{3}}}$, which is similar to the interlayer coupling constant found for different 2D vdw materials such as, $\mathrm{K_C = 8.62\times10^{19} {Kg}/{m^{3}}}$ in $\mathrm{MoS_2}$, \cite{zhao2013interlayer}, $\mathrm{K_C = 9.1\times10^{19} {Kg}/{m^{3}}}$ in $\mathrm{WS_2}$ \cite{yang2017excitation}, $\mathrm{K_C = 8.7\times10^{19} {Kg}/{m^{3}}}$ in $\mathrm{NbSe_2}$ \cite{he2016interlayer} etc.
The good fit of LCM with the experimental data implies that the interlayer vibrations in CGT are primarily governed by nearest-neighbor interactions, and that the interlayer coupling constant remains nearly independent of layer thickness—consistent with the fundamental assumptions of the LCM \cite{kittel2018introduction}.

\begin{figure}
	\centering
	\includegraphics[width=1\linewidth]{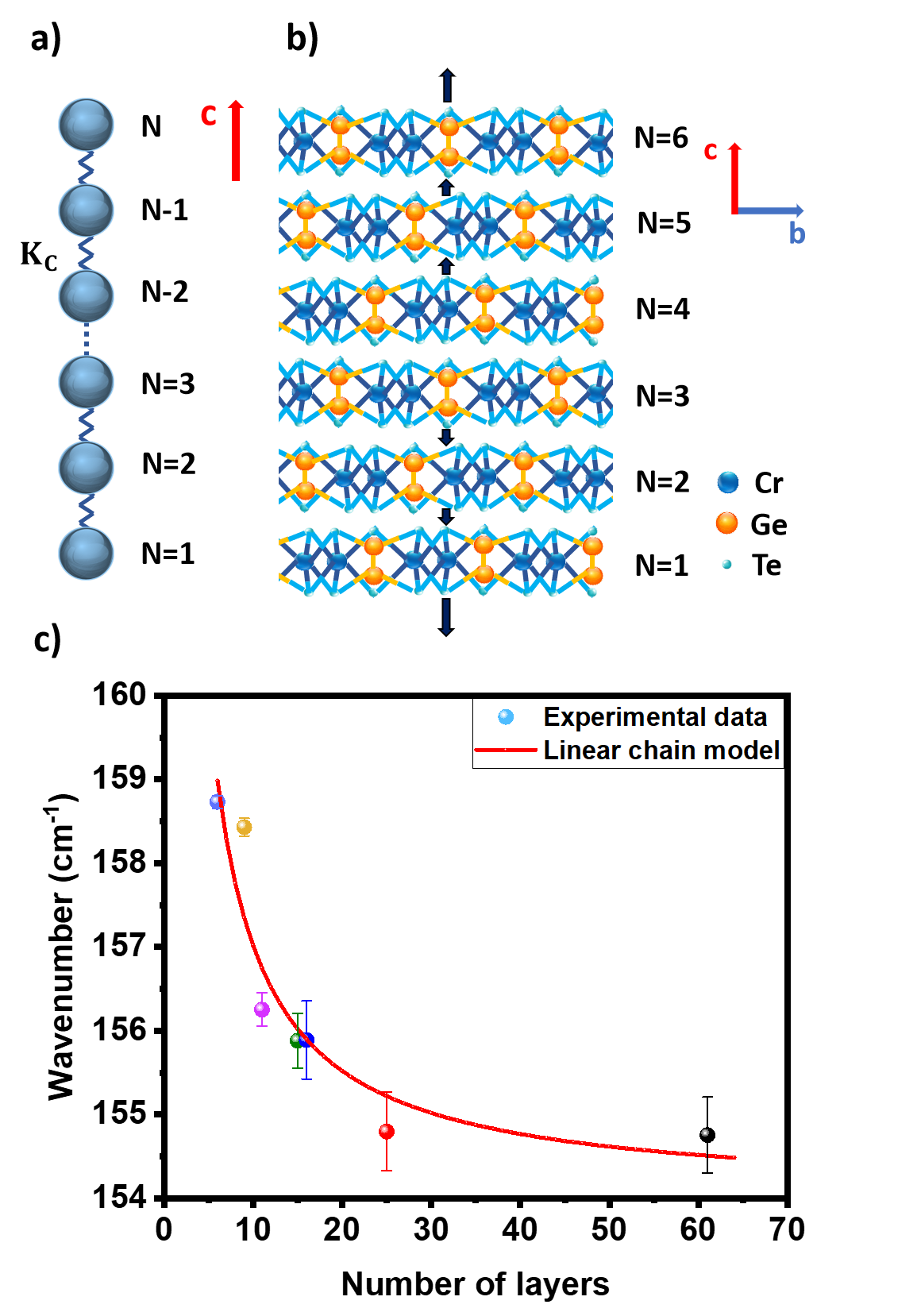}
	\caption{\label{fig:frog} a) Schematic of linear chain model where each layer is considered as a ball connected by a spring with interlayer coupling constant $\mathrm{K_c}$, b) Side view of the six layers of CGT, c) The experimental data points are fitted with LCM as indicated by the red line.}
\end{figure}

\section{Conclusion \\}

In conclusion, we have identified the layer breathing vibrational Raman modes in few-layer CGT around $158.73  {cm^{-1}}$ for S1, that arises from out-of-plane collective oscillations between adjacent layers. The symmetry analysis confirms that CGT belongs to the non-polar class of layered van der Waals materials, consistent with its centrosymmetric crystal structure. The layer-dependent evolution of the LBM frequency provides direct insight into the interlayer coupling strength, which we quantitatively evaluated using the Linear Chain Model (LCM) to extract the interlayer force constant ($\mathrm{K_c}$). The excellent agreement between the experimental data and the LCM fit confirms that the interlayer interactions in CGT are governed by van der Waals coupling.\\
These findings establish Raman spectroscopy as a powerful tool for probing thickness dependent interlayer vibrational mode, interlayer lattice dynamics, and symmetry-driven phonon behavior in 2D magnetic semiconductors. 

\section{Acknowledgments\\}
NC gratefully acknowledges financial support from the Department of Science and Technology, Govt. of India for INSPIRE Fellowship (IF22051). The authors also thank the Department of Physics, Indian Institute of Technology Delhi, for providing access to the Raman spectroscopy and atomic force microscopy (AFM) facility used in this work. PD acknowledges the funding support from Science and Engineering Research Board (SERB) of Govt. of India through grant no. SPR/2021/000762.

\section{Reference\\}
\bibliographystyle{unsrt}
\bibliography{aipsamp}
\end{document}